\documentclass[
  ,final            % use final for the camera ready runs
  ,numberedheadings % uncomment this option for numbered sections
  ]
  {aipproc}

\layoutstyle{8x11double}

\begin{document}

\title{Conference Summary}


\author{Brian W. O'Shea}{
  address={Theoretical Astrophysics Group, Los Alamos National Laboratory, Los Alamos, NM 87545; bwoshea@lanl.gov}
}

\author{Christopher F. McKee}{
  address={Departments of Physics and Astronomy, University of California, Berkeley, CA 94720; cmckee@astro.berkeley.edu}
}

\author{Alexander Heger}{
  address={Theoretical Astrophysics Group, Los Alamos National Laboratory, Los Alamos, NM 87545; aheger@lanl.gov}
}

\author{Tom Abel}{
  address={Kavli Institute for Particle Astrophysics and Cosmology, Stanford University,  Menlo Park, CA 94025;  tabel@slac.stanford.edu)}
}

\begin{abstract}
The understanding of the formation, life, and death of Population III stars, as well as the impact that these objects had on later generations of structure formation,
 is one of the foremost issues in modern
cosmological research and has been an active area of research during the past several
years.  
We summarize the results presented at ``First Stars III,'' a conference sponsored by Los Alamos National Laboratory, the Kavli Institute for Particle Astrophysics and Cosmology, and the Joint Institute for Nuclear Astrophysics.  This conference, the third in a series, took place in July 2007 at the La Fonda Hotel in
Santa Fe, New Mexico, U.S.A.
\end{abstract}

\maketitle

%%%%%%%%%%%%%%%%%%%%%%%%%%%%%%%%%%%%%%%%%%%%
%% MAINMATTER
%%%%%%%%%%%%%%%%%%%%%%%%%%%%%%%%%%%%%%%%%%%%

%%%%%%%%%%%%%%%%%%%%%%%%%%%%%%%%%%%%%%%%%%%%%%%%%%%%%%%%%%%%%%%%%%%%%%%%%%%%%
\section{Introduction}

In July 2007, more than 130 international researchers met at 
the La Fonda Hotel in
Santa Fe, New Mexico, U.S.A. to discuss the formation, life, and death of zero-metallicity (Population III) 
and very low metallicity stars,
as well as the impact that these objects had on later generations
of stars and on structure formation.  This field has made significant theoretical
and observational advances since First Stars I, which was a MPA/ESO/MPE Joint
conference held in 1999 in Garching b.\ M\"unchen, Germany, and
First Stars II, which was held 
at The Pennsylvania State University in State College, Pennsylvania, U.S.A., in 2003.  Though major 
advances have been made, the understanding of Population III and
very low metallicity stars is still in its infancy, and many 
important questions remain unanswered.  

Donald Rumsfeld is known for many things, but his contribution
to the philosophy of science is less well recognized. In an oft-cited
speech he delivered on Feb 12, 2002, he
said:

\begin{quote}
There are known knowns. These are things we know that we know. There are known unknowns. That is to say, there are things that we know we don't know. But there are also unknown unknowns. There are things we don't know we don't know.
\end{quote}

\noindent
He applied this to intelligence data, but
one can describe our task as scientists is to uncover ``unknown unknowns" by pure
thought or by serendipitous observation or experiment, thereby converting them to
``known unknowns;" and to then use systematic observation, experiment and
theory to convert these to "known knowns." 

The study of the first stars is a new field.
There are almost no ``known knowns," except for the background cosmology
and an increasing set of data on abundances in very metal poor stars. There
are many ``known unknowns:" What is the nature of dark matter? What is
the strength of the magnetic field? What are the abundance and size
distribution of dust? What is the rate of mixing of metals into
primordial gas? Until recently, the effect of dark matter annihilation on
the first stars was an ``unknown unknown," but this has become a ``known
unknown" through the work of Freese and her collaborators.

We conjecture that the number of ``unknown unknowns"---i.e.,
the discovery potential---in a field is proportional to the ratio
of the number of ``known unknowns" to ``known knowns:"
\begin{equation}
UU \propto KU/KK.
\end{equation}
If so, the study of the First Stars is an excellent field for young people---as reflected
in the youth of the audience and the organizers (although not in the summarizer)!

In the following sections, we shall summarize 
the results presented at this conference, and raise some significant 
issues that have yet to be explored.
In the interests of conserving space, we discuss only  
the results presented during First Stars III, and direct interested
readers to individual contributions for more in-depth information and
citations to refereed papers.  In addition, we have made a 
concerted effort to refer only  to contributions in this proceedings,
and do so by the last name of one of the authors (typically the first) 
of each contribution.

%%%%%%%%%%%%%%%%%%%%%%%%%%%%%%%%%%%%%%%%%%%%%%%%%%%%%%%%%%%%%%%%%%%%%%%%%%%%%
\section{A proposed naming convention for primordial and metal-poor stars}

A significant point of confusion in the literature on Population III and low 
metallicity stars arises from the abundance of (often contradictory and/or confusing)
naming conventions.  Given that essentially all researchers involved in the field 
were present at First Stars III, a discussion of this subject took place and
the following naming convention has been proposed:

\textbf{Population III:}  This is a blanket term that  describes 
all stars of primordial composition (i.e., gas whose composition was
determined during Big Bang Nucleosynthesis and as a result is composed almost
entirely of hydrogen and helium), regardless of how, when, or 
where they formed.  It has been found that this term is too broad,
and thus the need for the existence of ``\emph{Population III.1}'' and
``\emph{Population III.2}'' stars, as described below.

\textbf{Population III.1:}  These are the true ``first generation''
stars of primordial composition, whose properties have been determined entirely by 
cosmological parameters and the process of cosmological structure
formation, and have not been significantly affected by previous star
formation.  Examples of Population III.1 star formation are shown in 
contributions by Norman, Turk, and Yoshida.

\textbf{Population III.2:}  These are ``second generation''
stars, which still have primordial composition.  Their
formation, however, has been significantly affected by previous generations
of star formation through the injection of kinetic energy, photodissociating 
or ionizing radiation, by cosmic rays, or by other as-yet-unsuggested
processes, which may change the mass range of
these stars.  Examples of Population III.2 star formation are discussed
in contributions by Ahn, Bromm, Bryan, Johnson, McGreer, Sato, Norman, Umemura and Whalen.

\textbf{Population II.5:}  This is the suggested term for
a possible class of stars with non-zero metal content 
 where the amount of metals  in the gas that the stars are
formed from would not be sufficient to affect the cooling properties
of the gas and thus the formation of the stars, 
but would play a non-negligible role in the star's 
main-sequence evolution.  An example of this class of object (as discussed by Meynet)
is a rapidly rotating, massive star with a metallicity of 
$\simeq 10^{-6} \mathrm{Z}_\odot$, which would have formed in a manner identical
to a star of primordial composition, but would experience 
enhanced mass loss relative to a Pop III star of otherwise identical properties
due to the small amount of metals in the star.

\textbf{Population II:}  These are stars whose metal content
exceeds the ``critical metallicity'' discussed by Bromm, Glover, Omukai,  
Shull, and B. Smith -- namely, the metallicity where the enhanced cooling 
properties of metal-enriched gas affects the star formation process, possibly
resulting in a different IMF.  

%%%%%%%%%%%%%%%%%%%%%%%%%%%%%%%%%%%%%%%%%%%%%%%%%%%%%%%%%%%%%%%%%%%%%%%%%%%%%
\section{Formation and IMF of stars at zero and low metallicity}

The ultimate goal of the theoretical and numerical work being performed
by many investigators is to gain a fundamental understanding of the Initial
Mass Function (IMF) of Population III (Pop III) stars.  This is also true of
research relating to star formation in the present-day 
universe --  Population III  star formation, however, is considered
 to be a more tractable problem, given the relative simplicity
of the physics involved.

In the absence of observations of Population III stars, we must rely
on theory and simulations. As discussed by Norman, both theory
and simulations predict that the first stars 
were massive, with an average mass that is much greater than stars
in the Milky Way.  This is consistent with 
the absence of observed zero-metallicity stars at the present day.
Simulations show that not all 
Population III stars form in exactly the same fashion, however, 
which suggests a broad
spread of masses.  In addition, there is no evidence for fragmentation
in extremely high-resolution simulations of Population III star formation, 
even when theory suggests that there should be 
(though note that Umemura and Suwa displayed preliminary
calculations that indicate 
fragmentation in an asymmetric runaway collapse of primordial gas).

Turk and Yoshida presented simulations of Population III.1 
star formation using different methods.  Yoshida used the
Gadget-2 smoothed-particle hydrodynamics code, while 
Turk used the ENZO adaptive mesh refinement code.  Both codes
have been improved in the past few years with the addition of
extended-precision arithmetic, particle splitting techniques (for SPH), 
updated chemistry, approximations for cooling
that take into account the non-negligible optical depths at high
densities, and modifications to the ideal gas law equation of 
state at extremely high density.
These improvements allow both codes to simulate 
large ($\sim$~Mpc) volumes of the universe while at the same time
following the collapse of gas to protostellar densities, with a 
current maximum density of $n_H \simeq 10^{21}$~cm$^{-3}$! 
These fundamentally different methods agree quite well to
sub-parsec scales -- there is no evidence for fragmentation
in the collapsing primordial cloud cores, and the inferred
accretion rates of gas flowing onto the evolving protostellar
core are extremely high, peaking at $\dot{M} \simeq 10^{-2}-
  10^{-1}\,\mathrm{M}_\odot/\mathrm{yr}$.  At smaller scales, some differences
are apparent in their calculations.  It is unclear, however, 
whether this is due to numerical issues or simply due to 
the use of different cosmological realizations.  

Glover 
discussed issues relating to uncertainties in the 3-body molecular
hydrogen formation reaction rates.  At the temperatures 
relevant to Population III star formation ($T< 1000\,\mathrm{K}$), 
theoretical calculations of these rates vary by $2-3$ orders of magnitude.  

Theuns showed that 
Pop III star formation in warm dark matter may occur in a very
different manner than in a CDM universe, possibly leading to
filamentary fragmentation and the formation of intermediate-mass
primordial stars.  Freese presented results suggesting that if
neutralinos comprise a significant fraction of dark matter, their
annihilation could possibly overwhelm any cooling mechanism in high-density
primordial gas, and may result in a ``dark star'' -- a massive object powered
by dark matter annihilation instead of nuclear fusion.  Zhao and Xu both showed
preliminary results from AMR simulations including the effects of magnetic
fields and their production via the Biermann battery 
in cosmological simulations, which may have significant
effects on the formation of Pop III stars at small mass and
spatial scales.

An important issue with the presented simulations of
Population III star formation is that they are fundamentally
Courant-limited at high densities, so that the simulations
inevitably grind to a halt as the collapse proceeds.  
In the absence of a means to avoid this 
(e.g., sink particles), understanding of the shutoff of accretion onto
primordial protostars can only be obtained analytically.  To this end,
Tan presented analytic and semi-analytic calculations of the evolution
of the inner regions of primordial halos that follow the growth and 
evolution
of the protostar, formation of an accretion disk, and feedback 
from the evolving protostar and main-sequence Population III star.
He showed that the effectiveness of feedback processes in halting
accretion onto the star (and thus determining its final mass) depend
on core rotation and on the rate of accretion of gas onto the disk. 
Estimates of accretion rates from high-resolution cosmological
 simulations suggest stars with
masses of $60-400\,\mathrm{M}_\odot$, with the main-sequence 
stellar mass in the fiducial
case being $\simeq 160\,\mathrm{M}_\odot$.

Some aspects of Population III.2 star formation were 
also discussed.  Bromm (Greif et al.), Bryan, Yoshida, 
and McGreer  presented results showing the formation of 
primordial stars in regions of pre-ionized gas, using two different
methods.  All agree that significant amounts of HD form in these
regions, and result in a rapid cooling of gas down to the temperature
of the CMB at the redshift of formation.  This lower temperature
directly translates into a reduced accretion rate onto the protostellar
core, suggesting that the resulting Population III.2 stars will
be less massive than their Pop III.1 counterparts.  Bromm (Greif et al.) also
showed that the presence of a cosmic ray background can
cause the formation of significant amounts of HD in high-density
primordial gas, again lowering gas temperature to that of the CMB
and reducing the accretion rate.  This is not to say that all Population
III.2 star-forming regions are characterized by reduced accretion
rates:  Norman showed results from simulations of Pop III stars
forming in the presence of a molecular hydrogen photodissociating 
(Lyman-Werner) background, which results in higher overall halo
temperatures and higher inferred accretion rates onto the protostellar
core.  

The transition between Population III and Population II star formation
was discussed at length during the conference.  Bromm (Greif et al.) suggested the 
existence of a ``critical metallicity,'' or Z$_{\mathrm crit}$,  
where metal line cooling dominates over molecular hydrogen
cooling at low temperature (and hence presumably where the stellar IMF will
transition from being top-heavy to a Salpeter-type function with a significantly
lower mean mass), at $10^{-4}<Z_{\mathrm{crit}}/\mathrm{Z}_\odot <10^{-3}$.
 Glover and Shull both suggest that Z$_{\mathrm crit}$ may depend strongly on
environment.  Omukai showed that dust is an extremely effective coolant
compared to gas-phase metals, and that the presence of a small amount
of dust can radically lower the critical metallicity.  The degree with 
which this 
takes place depends strongly on the abundance,
type and size of the 
dust grains, and is highly uncertain.  
B. Smith used highly-resolved adaptive mesh refinement calculations to examine
the fragmentation of metal-enriched gas at small scales and observed
that there is not a clear relationship between metallicity and fragmentation
scale -- rather, the fragment scale is determined by the density of the
gas when it reaches the temperature of the CMB.  In this way,
higher-metallicity gas may reach the CMB temperature at lower densities
than lower-metallicity gas, resulting in larger overall clumps.
Jappsen performed SPH simulations of metal-enriched gas collapsing 
in an ionized halo.  She showed that, for densities $n_{\mathrm{H}} <
  10^4\,$cm$^{-3}$,
the evolution of density and temperature are not changed by metallicity
for $Z < 0.1\,\mathrm{Z}_\odot$, because H$_2$ is the dominant coolant rather than
metal fine structure lines.  In addition, she does not find evidence 
in her calculations for the
``critical metallicity'' threshold proposed by Bromm (Greif et al.)  Clark, however, used
high-density SPH simulations to show that fragmentation can easily occur
in the high-density, dust-dominated regime at metallicities at or below
$Z = 10^{-5}\,\mathrm{Z}_\odot$.

%%%%%%%%%%%%%%%%%%%%%%%%%%%%%%%%%%%%%%%%%%%%%%%%%%%%%%%%%%%%%%%%%%%%%%%%%%%%%
\section{Searches for Population III and very low metallicity stars and
observed abundance patterns}

\subsection{The search for Population III stars}

To date, no stars of primordial composition have been directly detected, 
either in our galaxy or in the distant universe.  This may be because these
stars are very massive and thus short-lived, or because surveys are looking
 in the wrong places.  If these objects still exist, how can they
be detected?  And, have we already indirectly detected the traces of Population III
stars?  

If, as Tan suggests, Population III.1 stars have masses of on the order of
$100-400\,\mathrm{M}_\odot$, some of these stars may explode as pair instability 
supernovae (PISN), as described by Woosley.  These objects would have unusual nucleosynthetic 
patterns, which have not yet been observed in abundance ratio measurements
of galactic halo stars.  This lack of evidence sets strong upper limits on on the 
number of 
primordial stars in this mass range.  

N. Smith presented observations of
SN2006gy, a Type IIn supernova in NGC1260 (a S0/Sa galaxy located approximately
$73\, $Mpc away).  This was the brightest supernova known at the time of the 
conference, and the light curve 
and inferred expansion velocity are inconsistent with other Type II supernovae
and cannot be explained by the interaction of a standard Type II supernova
with a circumstellar medium.  The observed light curve is  consistent with
results presented by Kasen showing theoretical predictions of pair
production supernova model light curves and spectra, lending hope that
Population III PISN exist.  In principle, a 
pair instability supernovae at $z \sim 20$ would be bright enough at 
peak luminosity to be observable with JWST,
though the time dilation of the light curve could make detection extremely 
difficult.  This would be much less of a problem if, as suggested by Schneider, 
a non-negligible fraction of stars forming at z~$\sim 3$ may be primordial.
In addition, if a significant fraction of Population III stars are GRB 
progenitors (as suggested by Yoon), JWST followups of SWIFT-detected
gamma ray bursts may yield direct observations of Population III supernovae.
Another model
for SN2006gy was presented by Woosley, a pulsational pair
instability supernova of about $110\,\mathrm{M}_\odot$.  In this
scenario the supernova-like ejecta of subsequent pulses run into
each other at large distance and convert their kinetic energy at
almost 100\,\% efficiency into radiation, making for a very bright
display.  Generally, the mass range for pulsational pair instability
in non-rotation primordial composition (Pop III) stars is about
$100-140\,\mathrm{M}_\odot$.

Examination of cosmic infrared background (CIB) fluctuations may yield indirect
detections of Population III stars.  Kashlinsky presented results from
observations using the Spitzer Space Telescope,
arguing that
source-subtracted IRAC images contain significant CIB fluctuations that are in excess
of the near-infrared background expected from resolved galaxy populations.  These
fluctuations appear to come from clustered sources that do not correlate with
Hubble ACS source catalog maps of the same field.  He claimed 
that this implies that the CIB 
fluctuations come from dim populations at high ($z > 8$) redshifts, with
much lower mass-to-light ratios than current galaxies, and that have a projected
number density of $5-10$ sources/arcsec$^2$ (within the confusion limit
of present-day instruments, but resolvable by JWST), and claimed that this 
source population is high-redshift primordial stars.  Thompson, on the
other hand, claimed that the purported near-infrared background excess is due
to improper modeling and subtraction of zodiacal light, and that 
the cosmic infrared background
fluctuations detected by Kashlinsky are mainly due to low-redshift 
($0.5 < z < 1.5$) galaxies and are inconsistent with galaxies at $z > 10$.
Fernandez showed that metal-free stars are not the only possible source of the
 near-infrared background excess -- her models predict that stars with metals can
produce the same amount of diffuse radiation in the $1-2$ micron band as primordial
stars, because the average intensity in this waveband is determined by the efficiency
of nuclear burning in stars, which is not very sensitive to metallicity.

\subsection{The search for low metallicity stars}

Metal-poor galactic halo stars are of great interest to the 
community that studies structure formation in the early universe
for several reasons.  Extremely metal-poor stars are 
 possibly fossil
records of the heavy element abundances produced in a single Population III
supernova.  The shape of the low-metallicity
tail of the metallicity distribution function (MDF) has the 
potential to probe epochs of star formation in the early galaxy,
and the change of the MDF with distance from the center
of the galaxy may provide useful clues as to the assembly history
of the Milky Way.  In addition, the discovery of objects with
enhancements of combinations of various $\alpha$-, s-, and r-process elements 
can probe rare events in the formation history of our galaxy.
Ferrara and Salvadori, however, argue that almost no true
``second generation'' stars (objects that have been
enriched by a single Population III supernova) should be
observable at $z=0$.

The discovery of metal-poor galactic halo stars and spectroscopic 
estimates of their elemental abundances has become a major
industry in the past several years.  The SDSS
SEGUE project (as discussed by Beers) is already underway, and has
discovered more than 2400 stars with [Fe/H]~$< -2$ and dozens with 
[Fe/H]~$< -3$.  Over the course
of the project lifetime, SEGUE is projected to find roughly 20,000 stars
with [Fe/H]~$< -2$, and approximately 2000 with [Fe/H]~$< -3$.
The 0Z Project (as summarized by Cohen) describes the discovery of
more than 1500 candidate extremely metal poor stars from the Hamburg/ESO
Survey, with some stars having extremely peculiar abundance ratios.
The ESO LAMOST Survey and Southern Sky Survey (described by Christlieb),
will begin production in the near future, will cover 
significantly larger areas of the sky than SEGUE (to a similar
magnitude) and will, as a result, find thousands of stars with
metallicities of [Fe/H]~$< -3$ in the next few years. 

Given the large number of extremely metal-poor stars that have been
observed, and the even larger number of stars that should be found
by surveys that are currently underway, a question arises -- how 
much can we trust the elemental abundances measured in these stars?
This is a crucial issue when one is interested in comparing the 
results from simulations (shown by Nomoto, Woosley, and others) with 
observationally-determined stellar abundances.
Sneden discussed some aspects of this important issue, and demonstrated 
that correct measurement of abundances depends strongly on accurate
determination of the star's effective surface temperature, and that
abundances derived from a small number of strong lines may be
unreliable.  In addition, laboratory data on transition probabilities
of alpha elements is often contradictory or completely lacking, and
that more accurate measurement of the properties of iron-peak elements is necessary
in order to accurately interpret abundance data.  Venn suggests that stars with the lowest
[Fe/H] abundances (that have [Fe/H]~$< -5$), which seem to be chemically
peculiar, may actually appear to be that way due to dust-gas
separation.  If this is true, stars with $-4 <$~[Fe/H]~$< -3$
may actually sample metal enrichment from single Population
III stars.  She also suggests searching nearby dwarf galaxies for
single pollution event stars in addition to the halo of our galaxy.

\subsection{Observed abundance patterns in low-metallicity stars}

Much can be inferred from the observed abundance patterns of extremely 
metal-poor galactic halo stars, though the concerns discussed in the 
previous section must be kept in mind when interpreting these results.

Frebel presented an update on the abundance of HE 1327-2326 using spectra
from the VLT, confirming its extremely low iron abundance and carbon 
enhancement [FeI/H] $= -5.7$ or $-5.5$, 
depending on assumptions, and [FeII/H] $< -5.4$, and [C/Fe]$_{\mathrm{LTE}} = 3.28$ using
a 3D model atmosphere correction).  This work was elaborated upon by Korn, who showed
that HE 1327-2326 is a subgiant and not a main-sequence star, and that atomic
diffusion may have significantly altered the surface abundances of the star
(though not enough to fully explain the puzzling non-detection of lithium). 
Frebel also presented work using observations of
carbon and oxygen-enhanced metal poor stars to examine 
necessary conditions for forming low-mass stars in the early universe through cooling 
via fine-structure lines, and showed evidence for a ``transition metallicity'' where
the sum of the carbon and oxygen abundance is approximately $10^{-3.5}$ the sum of
the solar abundances for these elements, which supports the Bromm idea of a ``critical
metallicity.''

Johnson argued that carbon-enhanced metal poor (CEMP) stars are created by the pollution of
a low-mass star by a companion asymptotic giant branch (AGB)
 star.  She suggested that the observed [C/N]
ratios imply that a large number of primordial stars with masses of $2 - 3$~M$_\odot$
were created.  This result was supported by Schuler, who presented observations of
fluorine in a CEMP halo star.  AGB stars are believed
to be prodigious producers of both carbon and fluorine, suggesting
that AGB stars may be the source of the observed abundance patterns 
in at least some CEMP stars.  Both of these observational results are supplemented
by Pols, who presented simulations of binary stellar evolution showing that 
the observed CEMP abundance ratios can be explained by binary pollution from
AGB stars, and that thermohaline mixing is important in the metal-poor companion star, and
by Husti, who presents similar results.
This latter effect is examined in more detail by Bisterzo, who presented numerical
experiments detailing the effects of thermohaline mixing on nucleosynthetic yields
resulting from neutron-capture nucleosynthesis.  These calculations show that 
the yields for CEMP stars can be well-matched when thermohaline mixing is included. 
Lucatello showed results from the HERES survey suggesting that there
are different classes of CEMP stars, based on the presence or absence of 
r- and s-process
elements.  These stars might be enriched by different types of stars, or via different
mechanisms, and the monitoring of stellar radial velocities may give some clues to
their different origins.   Rossi showed that estimates of carbon abundances in CEMPs 
may be affected
by calibration problems, and presented a set of refined estimates of carbon 
abundances for a large sample of CEMPs.

de Mink presented results from binary evolution at low 
metallicity.  She showed that binaries at low metallicity experience mass transfer
in a very different way than solar-metallicity stars do, which may have implications
for the metallicity dependence of the formation rate of various objects through binary
evolution channels.  She also showed that low-metallicity binaries can experience much
higher rates of accretion before reaching a given size compared to solar-metallicity
stars, suggesting that fewer low-metallicity binaries come into contact (and thus merge)
during rapid mass transfer.

Masseron presented results from the high-resolution 
chemical analysis of a large sample of CEMP stars, and showed that these stars
naturally split into two groups: stars enriched only in s-process elements, and those
enriched by both r- and s-process elements.  The former group is well explained by
AGB mass transfer, and Masseron suggests that the latter type of star can be explained
by a primordial companion with a mass of $8-10\,\mathrm{M}_\odot$.  Tumlinson argued that the 
two observed hyper metal poor stars were most likely formed in mass-transfer 
binaries with a top-heavy IMF, and that the high frequency of CEMP stars at 
[Fe/H]~$< -2$, and gradients with metallicity and location, suggest that the 
CMB sets the typical mass of early Pop II stars independent of metallicity.
Komiya also suggested that the IMF of extremely metal-poor stars is top heavy,
and that this is consistent with the observed metallicity distribution function
in metal-poor halo stars.

Sobeck presented results from a study of copper abundances in 50 metal-poor halo
stars using the VLT UVES spectrograph.  There appears to be a 
deficit of copper in these stars, implying the
reduced extent of weak s-process in massive stars at low [Fe/H] or the delayed
production from Type Ia supernovae.  Boesgaard discussed beryllium abundances in a sample
of 51 metal-poor halo dwarf stars that have been examined at 
high resolution and signal-to-noise using the Keck HIRES and Subaru HDS spectrographs.
There is no evidence for a beryllium ``plateau'' -- rather, [Be/Fe] remains constant
with changing [Fe/H] -- and also no evidence for a difference in Be vs. [O/Fe] in 
stars that are found in the halo versus those observed in the thick galactic disk,
contrary to previous lower-resolution work.

Cowan presented results from HST and ground-based studies of galactic halo stars,
and argued that at the time these stars formed the galaxy was chemically unmixed
and inhomogeneous in r-process elements, but not in $\alpha$-elements, and suggests 
different environments for the synthesis of these elements.  There is evidence
for increasing contribution of the s-process with metallicity (and thus galactic
age).  Krugler presented results from autoMOOG analysis of over 6500 stars from 
SDSS-I and SEGUE that have estimated metallicities of [Fe/H] $< -2$ and
effective temperatures of $4500\,\mathrm{K} < T_{\mathrm{eff}}
  < 7000\,\mathrm{K}$.  This technique
produces estimates of [Fe/H] and [Ce/Fe] (or upper limits on these quantities).
Roederer derived isotopic fractions of europium, samarium and neodymium in two metal-poor
giants with apparently different nucleosynthetic histories, extending the examination 
of the neutron-capture origin of multiple rare-earth elements in metal poor stars to 
the isotopic level for the first time.  The results suggest an r-process
origin for the rare earth elements in one of the stars, and cannot distinguish between an r-
or s-process origin in the other star.  This is an important step, however, towards being able
to compare nucleosynthetic predictions for the s- and r-process using this technique.

Lithium is an element of great cosmological significance.  There are two stable 
isotopes -- $^6$Li and $^7$Li -- with $^7$Li being formed during the epoch of Big
Bang Nucleosynthesis (BBN), but not $^6$Li.  There is an apparent discrepancy between the
measured amounts of $^7$Li and the predicted amounts based on recent determinations
of the baryon-to-photon ratio, with the observed amount of $^7$Li being too low.  
Asplund pointed out that there are actually two cosmological lithium problems -- 
the observed abundance of $^7$Li is
inconsistent with predictions from standard BBN, and $^6$Li is apparently inconsistent
with galactic cosmic ray production.
He suggests a range of possible mechanisms to
cause these inconsistencies, but cautions that they state inconsistencies 
are based on extremely challenging observations and analysis, and should be
taken with a grain of salt.  
Perez presented observations of two metal-poor stars
 suggesting isotopic abundance ratios of $^6$Li/$^7$Li $\simeq 0.04-0.05$.  In one
case this is similar to results found in the literature for the same star, and in
the other case her derived value is significantly higher.  Perez suggested that the derived
abundance ratio is extremely sensitive to parameters used in the analysis.
Sbordone also found this sensitivity, and showed that effective temperature scales
for extremely low metallicity stars are still poorly calibrated, leading to inaccurate
determination of lithium isotopic abundance ratios.

%%%%%%%%%%%%%%%%%%%%%%%%%%%%%%%%%%%%%%%%%%%%%%%%%%%%%%%%%%%%%%%%%%%%%%%%%%%%%
\section{Stellar evolution, explosions and nucleosynthesis at zero and very
low metallicities}

\subsection{Stellar Evolution and Explosions at Very Low Metallicities}

In his review talk, Woosley made several points.  He claims that the known
abundances in low metallicity stars can be fit by ``ordinary'' supernovae in the
$10 - 100$ M$_\odot$ range, and that there is no need for hypernovae or pair
instability supernovae.  He argues that the favored Pop III stellar masses based on
nucleosynthetic results are $10-20\,\mathrm{M}_\odot$, with explosion energies of
roughly $10^{51}\,\mathrm{ergs}$, and little mixing.  He also suggested that metal-deficient
stars will produce many more black holes than their metal-rich counterparts, with
masses as high as $40\,\mathrm{M}_\odot$.  Finally, he argued that the pulsational
pair instability can give a wide range of light curves, from faintest to
the brightest observed supernovae.  Yoon presented some results based on
simulations of massive stars at very low metallicity, including rotation and
binary interaction.  He claims that the effects of rotation are particular
important for the evolution of massive stars at low metallicity, and that a 
large fraction of these stars may produce GRBs or hypernovae with unique
nucleosynthetic features.  Additionally, he suggested that a significant
fraction of massive stars may be Wolf-Rayet stars at very low metallicity,
and that stars of this type in close binaries may also produce GRBs.  Meynet
also spoke about the evolution of Pop III and very metal poor stars, and
stated that the effects of rotation on Pop III stars is significant, but
less extreme than in very metal poor stars.  He also suggested that a tiny
amount of metals (on the order of $Z = 10^{-8}\,\mathrm{Z}_\odot$) may make a big
difference, and that rotation is a key parameter for very metal
poor stars.  Ekstrom argued that under certain conditions it may be possible
for very massive primordial stars to avoid pair-instability supernovae with
the help of two effects of rotation: anisotropic winds and magnetic fields. 
This will happen even with the assumption of reasonable initial equatorial
velocities.  Chiappini presented further results regarding the impact of 
stellar rotation on chemical enrichment at low metallicity, showing that
fast stellar rotation at low metallicities is the only thing that can
explain the observed abundances in metal-poor halo stars in the 
absence of AGB binary mass transfer.  Lau
presented simulation results from explosions of 
intermediate-mass, zero-metallicity
stars, and showed that the stellar envelopes of these stars will be enriched
by nitrogen, and that there is no s-process enrichment, owing to the lack of
a third dredge-up.  Gil-Pons presented results from simulations that examine
the effects of overshooting in the evolution of intermediate-mass stars
that agrees well with Lao's result. 
Brott discussed the efficiency of rotational mixing in massive stars, 
and finds that while internal magnetic fields are necessary to understand
angular momentum transport (and thus rotational behavior), the corresponding
chemical mixing must be neglected in order to reproduce observations. She
also showed that for low metallicity stars, detailed initial abundances are
of primary importance, since solar-scaled abundances may result in significant
calibration errors.  Tsuruta presented results from simulations of the evolution
of very massive ($500\,\mathrm{M}_\odot$ and $1000\,\mathrm{M}_\odot$) Population III stars, and finds that
though these stars experience the pair instability, they eventually undergo
core collapse and create black holes that, in the more massive star, may be
up to $500\,\mathrm{M}_\odot$.

Pols presented theoretical results pertaining to carbon-enhanced metal-poor (CEMP)
stars, arguing that these stars are in binary systems and have been
polluted by a former AGB companion, and also suggesting that it is important
to study nucleosynthesis together with binary evolution.  Suda 
discussed the stellar evolution of low- and intermediate-mass extremely metal
poor stars, and suggested that these stars may be responsible for CEMP stars.
Ludwig discussed hydrodynamical model atmospheres of metal-poor stars,
and showed that metal-poor stellar atmospheres are prone to exhibiting substantial
deviations from radiative equilibrium.  He also suggested that large abundance corrections
may have to be made to take into account assumptions made in 1D atmosphere
calculations.  In addition, the three-dimensional
effects on the effective atmospheric temperature from Balmer lines show a complex
pattern, further complicating analysis.  van Marle showed results from simulations
of continuum-driven winds from primordial stars, and argued that continuum 
driving can produce strong mass loss from stars, even without metals.  
Krticka presented the possibility that hot Population III stars may experience
significant mass loss if their atmospheres are enriched by CNO elements through
metal line-driven winds, with a range of effects depending on the metallicity
of the atmosphere.  Muijres presented results from simulations exploring the effect
of clumping on predictions of mass-loss rate of early-type stars, and suggests that
the difference between theoretically expected and empirically derived mass-loss rates
may be due to inhomogeneities.  She predicts that clumping leads to a higher mass-loss
rate, with only modest clumping factors required to match observed values of mass-loss
rates in massive stars.  Onifer showed attempts to calculate the mass-loss rate of
a Population III Wolf-Rayet star using a modified version of the CAK approximation.
He showed that even a star with zero initial metallicity will experience significant
mass loss due to radiation pressure on dredged-up nucleosynthetic products.

Church presented multidimensional simulations of  
 primordial supernovae, which investigate the effects of 
Rayleigh-Taylor-induced mixing and asymmetries in the explosion on the final
composition of the escaped gas.  She finds that for spherically-symmetric
explosions, mixing has little effect on the shells interior to oxygen, and that
some asymmetry is needed in the explosion in order for elements interior to 
oxygen to escape from the star.
Nozawa showed the evolution of dust grains that formed in Population III supernovae,
and argued that dust grain size and composition strongly affects their evolution. He
also demonstrated that small grains are preferentially destroyed in primordial supernova
remnants, with the maximum destruction mass varying as a function of the ambient
medium properties.  

\subsection{Nucleosynthesis at Zero and Very Low Metallicities}

Qian presented a review of nucleosynthesis of metal-free and metal-poor stars.  In
his talk, he made several points; he suggested that the origin of $^6$Li in 
low metallicity stars is still a puzzle; that the presence of nitrogen in 
low metallicity stars indicates rotation; and that the presence of lead in 
metal-poor binary members indicates that the s-process occurs at low metallicity.
He also argued that the observed abundances in metal-poor galactic halo
stars suggests standard supernovae rather than pair-instability
supernovae, but that there are contributions from both hypernovae and faint
supernovae.  Nomoto discussed nucleosynthesis in massive Population III 
stars, focusing on hypernovae and jet-induced explosions.  He showed that
explosions with large energy deposition rates will create gamma ray bursts with
hypernovae
and that their yields can explain the abundances of normal extremely metal-poor
stars, and that explosions with small energy deposition rates will be
observed as GRBs without a bright supernova, and can be responsible
for the formation of CEMP and hyper metal poor stars.
Kratz (Farouqi et al.) presented results exploring
nucleosynthesis modes in the high-entropy-wind scenario of Type II
supernovae, and showed that a superposition of several entropy components
can reproduce the overall solar system isotopic r-process residuals, as 
well as the more recent observations of elemental abundances of metal-poor,
r-process-rich halo stars.  Pignatari presented results of simulations
examining the weak s-process at low metallicity, and showed that the s-process
efficiency changes significantly as a function of both stellar mass and metallicity. 

Woodward and Herwig (combined proceedings) discussed nucleosynthesis and 
mixing in the first generation of low- and intermediate-mass stars, with 
a focus on studying entrainment at convective boundaries.  They point out
that one feature of stars at zero and low metallicity is that convective-reactive
mixing events are common, which is not true in stars of higher metallicity.
They show that 1D models have difficulty simulating these evolutionary
phases and the mixing that comes from them, and show 2D and 3D simulations
of this process.  Cristallo presented work on the evolution and nucleosynthesis
of low-mass metal-poor AGB models with carbon- and nitrogen-enhanced opacities (to
address observed carbon and nitrogen-rich metal poor stars), and shows that the
new opacities can cause significant changes in the chemical and physical
evolution of the stars, and may cause non-negligible changes in the amount of
carbon, nitrogen, and s-process elements created.  Campbell discussed the structural and
nucleosynthetic evolution of metal-poor and metal-free low and intermediate-mass
stars from ZAMS to the end of the AGB phase, and showed that 
many of these stars experience violent evolutionary episodes
that are not seen at higher metallicities.  These episodes may be coupled
with strong mixing, causing surface pollution.
Surman presented simulations results of nucleosynthesis in outflows from Kerr black hole
accretion disks.  She suggests that these accretion disks may be important contributors 
to the nuclear abundances in the oldest stars, particularly for rare species or those
not uniformly observed.

%%%%%%%%%%%%%%%%%%%%%%%%%%%%%%%%%%%%%%%%%%%%%%%%%%%%%%%%%%%%%%%%%%%%%%%%%%%%%
\section{Early feedback processes: influence on structure formation, reionization,
and IGM abundance patterns}

The first generations of stars are potentially prodigious sources of
mechanical, chemical, and radiative feedback, and may contribute significantly
to the enrichment and reionization of the intergalactic medium.  Reed showed
that high-redshift halos are strongly biased, particularly at small scales -- this effect
differs from low-redshift bias, and is a function of mass.  The
clustering observed will significantly affect the statistical properties of 
feedback at high redshifts.  Ciardi
argued that feedback from Population III stars is generally not as 
efficient as one might expect
from energetics estimates, and that objects with masses of $M> 10^{7-8}\,\mathrm{M}_\odot$
are not greatly affected by feedback.  
Wise presented AMR simulations modeling the formation of a significant
number of Pop III stars in a single volume, and showed that dynamical
feedback from HII regions and supernovae can expel most of the gas in
first-generation halos and lead to low baryon fractions in subsequent
star-forming halos.  In addition, he showed 
that metals are well-mixed
within dwarf galaxies, and that ejecta from Pop III supernovae will provide
a maximum metallicity of $\sim 10^{-4}\,\mathrm{Z}_\odot$. Greif presented an SPH calculation
of a Pop III pair-instability supernova expanding into a HII region formed by 
the progenitor star, and found 
similar results -- the supernova completely disrupts
the host halo and expands out to a significant fraction of the size of the HII
region, ultimately polluting $\simeq 2.5 \times 10^5\,\mathrm{M}_\odot$ of gas with metals.
Nagakura presented 1D, spherically-symmetric simulations of the evolution of supernova
remnants in the early universe, paying particular attention to the thermal and chemical
evolution of the expanding dense shell of gas.  He showed that at high redshift, 
regardless of metallicity,
 the minimum temperature of dense gas in the supernova remnant is limited by the
 CMB temperature, suggesting that fragmentation of the shell depends more critically on 
the density of the ambient medium and the supernova energy than on the metallicity
of the gas.  Nozawa discussed dust evolution in Pop III supernova remnants, and showed that 
the transport of dust within the evolving SNR depends strongly on the size and composition 
of the dust grains, and that small dust grains are preferentially destroyed in 
the remnant evolution.  Additionally, dust can be strongly segregated from metal-rich gas,
possibly explaining the abundance patterns of iron, magnesium and silicon in 
the lowest metallicity stars (assuming they were formed in the shells of Pop III supernova
remnants).

Whalen presented 2D radiation 
hydrodynamical simulations of the photoevaporation of minihalos by neighboring
Population III stars, and demonstrated that, when appropriate physics and
coordinate geometries are used, feedback from the I-fronts of neighboring
halos will largely be positive or neutral, and that the impinging radiation drives primordial
chemistry that is key to the hydrodynamics of the halo, making
multifrequency radiation transport necessary.  Umemura and Sato confirmed
this result with 3D SPH simulations including multigroup radiation
transport, and additionally suggested that  the
shielding due to H$_2$ created in the ionization front may allow the
formation of multiple stars in a single halo.  In contrast, Ahn 
presented work using a 1D Lagrangian radiation transport code that 
suggests that the result of an impinging I-front will be roughly neutral.
Bryan and McGreer showed that HD
cooling can play an important role in low-mass non-pre-ionized halos,
reducing accretion rates onto the protostellar cloud core.  Johnson 
(Greif et al.) showed that
primordial stars forming in relic HII regions would likely be protected from
radiative feedback from neighboring halos due to the large amounts of 
molecular hydrogen that form even at low densities due to the high residual
electron fraction.

Schneider presented a set of cosmological simulations of $5-10\,$Mpc boxes
that include a metallicity-dependent
star formation and feedback algorithm  algorithm.  These simulations
indicate that, due to inefficient metal enrichment, Population III
star formation may continue up to $z \sim 2.5$ in pockets of primordial
 gas.  The sites where Pop III stars form also change 
over time, moving from the centers of halos (at high redshifts) to 
the outskirts (at low redshifts).  In addition, she suggested that
only $1\,\%$ of metal-enriched gas has been polluted purely by Population
III supernovae, making nucleosynthetic signatures hard to identify. Pieri
performed simulations of comparable volumes that also take into account 
photoionization suppression of halo collapse.  He finds that radiative
feedback has a strong impact on mechanical feedback, and reduces metal
enrichment at essentially all overdensities.

Ricotti demonstrated that an X-ray background from massive Population III supernovae,
supernova remnants, and from accretion onto intermediate-mass black holes
may be an important source of ionization in the IGM, and that the topology
of this reionization will be ``outside-in,'' with voids being ionized first.  In
addition, redshifted X-rays will contribute to HI and HeII reionization in the
low-density IGM.  Ferrara presented a model for reionization that includes effects
from Pop III stars, Pop II stars, and AGN.  This model predicts that reionization
started at $z \sim 20$ by Population III stars and was 90\% complete by $z=7$,
and that quasars  dominate only reionization at z~$< 6$.
Additionally, the model predicts that more than $80\,\%$ of ionizing photons
responsible for reionization
were produced at $z \geq 7$ from halos with $M < 10^9\,\mathrm{M}_\odot$, 
implying
that the bulk of the reionization sources at high $z$ 
have not yet been observed.  This
model is consistent with the high-redshift QSO absorption line measurements
presented by Fan and with several other sets of observations, including
the observed number of Lyman-alpha emitters and damped Lyman-alpha systems.
Trac and Shin presented results from large volume, high-resolution dark matter-only 
simulations that have been post-processed to model the effects of 
radiative feedback from Population III and metal-enriched stars, and 
found results that are also in good agreement with Fan's observations
of the Gunn-Peterson trough in high-redshift quasar absorption lines,
and are also consistent with the most recent WMAP electron optical depth
measurements.  Iocco presented results showing that Pop III stars
will produce a diffuse high-energy neutrino background -- the strength of
this background, however, will fall below detectable thresholds for all
current and planned neutrino telescopes even with highly optimistic 
assumptions, ruling out the neutrino background as a useful diagnostic
tool for metal-free stars.  This was confirmed by Suwa, who also estimated
gravitational wave emission from the collapse of isolated Population III
stars, and suggested that the gravitational wave background from these
objects could be detectable by future gravitational wave interferometers
such as BBO or DECIGO.

%%%%%%%%%%%%%%%%%%%%%%%%%%%%%%%%%%%%%%%%%%%%%%%%%%%%%%%%%%%%%%%%%%%%%%%%%%%%%
\section{Compact objects at high redshift}

\subsection{Gamma-Ray Bursts and Quasars}

Rockefeller showed three-dimensional simulations of the collapse of
a massive Population III star, demonstrating that it is possible to
create a rapidly-accreting disk of gas around the central black hole, which
can then eject significant amounts of the stellar envelope.  Li used these
simulations as initial conditions for a three-dimensional MHD calculation
of the collapsing stellar core, showing that the star's magnetic energy can drive
an outgoing shock, which leads to an explosion.  In addition, the magnetic energy
is primarily in ``bubbles'' of gas that are dominated by magnetic energy,
and tend to evolve in a very inhomogeneous (and collimated) fashion.  This
has profound implications for Population III stars as gamma ray burst progenitors, and also for nucleosynthesis.

High-redshift gamma ray bursts have proven to
 be extremely useful tools for probing the ISM and IGM.  
Chen discussed using spectroscopy of gamma-ray burst afterglows
to perform direct, detailed studies of the ISM in the star-forming regions
of distant galaxies, which is impossible with quasars.  A large 
sample of GRB sightlines, coupled with moderate resolution afterglow
spectra and imaging of the host galaxies, will be key tools for understanding
the properties of the ISM in high-redshift galaxies.  Penprase showed
more results related to GRB afterglows, including highly detailed estimates
of the properties of star-forming regions in damped Lyman-alpha systems,
and demonstrated the promise of GRBs as probes of star-forming regions
in high-redshift galaxies.  Yonetoku showed a strong correlation between the
spectral peak energy of prompt GRBs and the peak luminosity, and used this
to estimate the possible redshifts for hundreds of BATSE GRBs of unknown redshift.
This analysis predicts that a large number of massive stars were formed in the
early universe, assuming Pop III stars form GRBs. 

Quasar absorption line studies are complementary tools to the examination
of GRB afterglows, 
in that they are very useful probes of the IGM at high redshift.
At present, measuring the Gunn-Peterson trough in QSO absorption line
spectra is one of the primary methods of constraining the end of the
epoch of reionization.  This was shown by Fan, who demonstrated that the
evolution of the ionization state of the IGM experienced a very
rapid change at $z \sim 6$, increasing by an order of magnitude
in $\Delta z \simeq 0.5$.  This implies that $z \sim 6$ marks
the end of the stage of inhomogeneous, overlapping bubbles of 
reionizing photons.  There is also evidence that $z \sim 6$ quasars
are quite old, suggesting a great deal of black hole buildup
and chemical enrichment at high redshift.  In addition, there
are far too few quasars at $z \geq 6$ to contribute 
significantly to reionization, suggesting that, unless the high-redshift
QSO luminosity function is very steep, the dominant
source of ionizing photons in the high-redshift universe is
stellar populations.

\subsection{Supermassive black hole progenitors}

Can Population III stars be supermassive black hole progenitors?  There
are many examples of $z \sim 6$ quasars, which are believed to be
black holes with masses $M_{\mathrm{BH}} \geq 10^6\,\mathrm{M}_\odot$ sitting
at the centers of massive galaxies.  Trenti argued that these
are extremely rare objects (one per~\ $\simeq
  0.5\,\mathrm{Gpc}^3$ at $z = 6$), and
that rare, extremely high redshift ($z \sim 40-50$) black hole seeds
can be QSO progenitors.  Alvarez, on the other hand, argued that high-redshift
black holes form in halos where the progenitor star has expelled the 
majority of the surrounding gas, initially suppressing accretion.  After gas
collects in the halo again, there will be strong self-regulation
of accretion by radiative feedback. As a result, black holes formed from 
Pop III stars will not accrete rapidly enough to be SMBH
progenitors (this also has implications for the X-ray ionization scenario
discussed by Ricotti)

Begelman suggested that it may be possible to form extremely 
massive ``quasistars'' by the direct collapse of gas in
cosmological halos, without a stellar precursor.  The
required precursor object could be generated by the formation of
halos with T$_{vir} > 10^4$~K, or possibly in the aftermath of a
large halo merger.  These objects would be radiation-dominated,
and would feature the formation of a central black hole, which then
becomes a source of energy that would create a convective envelope.
This object could cool via thermal neutrinos, allowing it
 to accrete at the Eddington limit of the envelope, rather than
the black hole itself.  This ``quasistar'' would have a low
effective surface temperature ($T_{\mathrm{eff}} \sim 4000\,\mathrm{K}$), 
would
have a lifetime on the order of $10^6$ years, and could potentially
be observed by JWST.

Colgate argued that supermassive black holes form naturally as a
result of cosmological structure formation -- high-entropy gas
creates $\sim 10^5\,\mathrm{M}_\odot$ stars, which collapse due to
relativistic instabilities and become the central point of a massive
accretion disk, which can then channel gas into the
black hole at super-Eddington rates, using a combination of self-gravity
instabilities and Rossby vortices to efficiently transport angular momentum
outward.
 
%%%%%%%%%%%%%%%%%%%%%%%%%%%%%%%%%%%%%%%%%%%%%%%%%%%%%%%%%%%%%%%%%%%%%%%%%%%%%
\section{Summary}

The properties of the first generations of stars, as well
as their effects on later epochs of cosmological structure formation,
have long been a topic of great interest to the astrophysical
community.  Progress in understanding these objects has been
significant and rapidly accelerating in the past few years, 
and many insights have been obtained by theory and by large-scale numerical
computations.  Progress has also been made observationally, in 
a wide variety of settings, and many significant questions have
been raised.  
The impressive array of new facilities and
surveys that are currently running or are planned for
the next few years, together with increasingly powerful
simulations,  should 
yield crucial new insights into
the earliest epochs of star formation in the universe. 

%%%%%%%%%%%%%%%%%%%%%%%%%%%%%%%%%%%%%%%%%%%%%%%%
%% BACKMATTER
%%%%%%%%%%%%%%%%%%%%%%%%%%%%%%%%%%%%%%%%%%%%%%%%

\begin{theacknowledgments}
This conference was generously supported by Los Alamos National Laboratory
through the Center for Space Science and Exploration, the Institute for Geophysics
and Planetary Physics, and the Theoretical Division.  In addition, support was 
provided by the New Mexico Consortium Institutes for Advanced Study,
the Kavli Institute for Particle
Astrophysics and Cosmology at Stanford University, and the Joint Institute for
Nuclear Astrophysics.
BWO and AH are supported under the auspices of the
National Nuclear Security Administration of the
U.S. Department of Energy at Los Alamos National
Laboratory under Contract No. DE-AC52-06NA25396.  BWO was 
supported by a LANL Director's Postdoctoral Fellowship (DOE LDRD grant 
20051325PRD4). AH was also supported by the DOE Program for Scientific Discovery
  through Advanced Computing (SciDAC; DE-FC02-01ER41176).
The research of CFM is supported by the National Science Foundation through
grants AST-0606831 and PHY05-51164 and by the Department of Energy through grant
DE-FC02-06ER41453.

\end{theacknowledgments}

\end{document}